\documentclass[twocolumn]{aastex63}
\usepackage[normalem]{ulem}  
\usepackage[T1]{fontenc}
\usepackage{ae,aecompl}
\usepackage{graphicx}
\usepackage{amsmath}
\usepackage{amssymb}
\usepackage{supertabular}
\usepackage{hyperref}
\usepackage{xcolor}
\usepackage{academicons}

\def\bea{\begin{eqnarray}} \def\eea{\end{eqnarray}}
\def\fm3{\;\text{fm}^{-3}}
\def\mev{\;\text{MeV}}
\def\gev{\;\text{GeV}}
\newcommand{\msun}{\,M_{\odot}}
\newcommand{\km}{\hbox{$\,{\rm km}$}}

\begin{document}

\title{Two-flavor color superconducting quark stars may not exist}

\author{Wen-Li Yuan}
\affiliation{School of Physics and State Key Laboratory of Nuclear Physics and Technology, Peking University, Beijing 100871, China}
\affiliation{Department of Physics, Nanjing University, Nanjing 210093, China}
\affiliation{Department of Astronomy, Xiamen University, Xiamen, Fujian 361005, China}

\author[0000-0001-9849-3656]{Ang Li}
\affiliation{Department of Astronomy, Xiamen University, Xiamen, Fujian 361005, China}

\correspondingauthor{Ang Li}
\email{liang@xmu.edu.cn}

\date{\today}

\begin{abstract}
Large uncertainties in the determinations of the equation of state of dense stellar matter allow the intriguing possibility that the bulk quark matter in beta equilibrium might be the true ground state of the matter at zero pressure. And quarks will form Cooper pairs very readily since the dominant interaction between quarks is attractive in some channels. As a result, quark matter will generically exhibit color superconductivity, with the favored pairing pattern at intermediately high densities being two-flavor pairing. In the light of several possible candidates for such self-bound quark stars, including the very low-mass central compact object in supernova remnant HESS J1731-347 reported recently, we carry out one field-theoretic model, the Nambu–Jona-Lasinio model, of investigation on the stability of beta-stable two-flavor color superconducting (2SC) phase of quark matter, nevertheless find no physically-allowed parameter space for the existence of 2SC quark stars. 
\end{abstract}

\keywords{
Compact objects (288); 
High energy astrophysics (739)
}

\section{Introduction} 

It is generally believed that the degree of freedom of dense matter is hadronic around the nuclear saturation density, $\rho_0 \sim 0.16 \fm3$, which undergoes a chiral phase transition or deconfinement phase transition at supranuclear densities to become quark matter. 
A recent analysis of the supernova remnant HESS J1731-347 suggested that the mass and radius of the central compact object within it might be $M=0.77^{+0.20}_{-0.17}\msun$ and $R=10.4^{+0.86}_{-0.78} \km$, respectively~\citep{2022NatAs...6.1444D}. 
The possible existence of ultra-low-mass compact stars has sparked interest in speculating about self-bound quark matter as the potential physical nature of pulsar-like objects~\citep{1971PhRvD...4.1601B,1984PhRvD..30..272W}.
For example, based on the mass-radius relations obtained from the MIT bag model 
and Nambu–Jona-Lasinio (NJL) model, as well as the constant speed of sound parametrization,
\citet{2022arXiv221107485D} demonstrated that this compact star might be a strange star,
while further indicating an existing astrophysical path that could lead to the formation of such an object.
Nonstrange quark stars compatible with the compact object HESS J1731-347 were studied in~\citep{2023PhRvD.107k4015R}. 
The study employed a developed resummation technique known as renormalization group optimized perturbation theory (RGOPT) to determine the equation of state (EOS) describing nonstrange cold quark matter at next-to-leading order.

Nevertheless, as quantum chromodynamics (QCD) predicts an attractive force between quarks in the color-antitriplet channel, it is expected that quarks near Fermi surfaces will form Cooper pairs, resulting in color superconductivity. 
And the color-flavor locked (CFL) phase is anticipated as the favored pairing pattern at asymptotic densities~~\citep{1998PhLB..422..247A,1999NuPhB.537..443A,2008RvMP...80.1455A}. 
Along this line, the CFL quark stars have been studied within MIT bag models in light of the compact object in the HESS J1731-347.~\citet{2023A&A...672L..11H} showed that candidate XMMU J173203.3-344518 might fit into a strange star scenario that could be
also consistent with heavier compact stars using the same quark EOS in the MIT bag model, and discussed a possible scenario where a small gap $(\Delta \leq 1 \mev)$ in the CFL phase would explain the high surface temperature of XMMU J173203.3-344518.
\citet{2023PhRvD.108f3010O} indicated that the parameter space could be further confined to the additional requirement for a high maximum mass $(M_{\rm{TOV}} \geq 2.6 \msun)$~\citep{2020ApJ...896L..44A}, without violating the conformal bound~\citep{2022PhRvL.129y2702F}. 

In this study, using the NJL-type model that incorporates quark-antiquark vector interaction and scalar diquark interaction at the microcosmic level, we aim to explore the stability of two-flavor color superconducting (2SC) quark matter, which is suggested to exist at intermediate densities associated with compact stars~\citep{2001ARNPS..51..131A,2008RvMP...80.1455A}.
We take into account the charge neutrality conditions that must be imposed for a stable bulk phase. 
Previous studies have investigated the consequences of maintaining electric and color neutrality in the color superconducting phase.
Existing research has primarily focused on fundamental aspects like the size of the pairing gap or the free energy of the system~\citep{2001PhRvL..86.3492R,2002PhRvD..66i4007S,2003PhRvD..67f5015H,2004PhRvD..69a4014M,2005PhRvD..72e4024S,2005PhRvD..72f5020B,2005PhRvD..72c4004R}.
They, however, have not questioned the absolute stability of bulk 2SC quark matter, where the energy per baryon should be lower than that of a $^{56}\rm Fe$ nucleus $E/A (p=0)=(56m_N-56\times8.8\mev)/56=930\mev$~\citep{1971PhRvD...4.1601B,1984PhRvD..30..272W}\footnote{Note that quark matter in gravity-bound neutron stars is not in absolute stability state and it only appears through a phase transition from hadronic matter at high densities~\cite[see more discussions in e.g.,][]{2020ApJ...904..103M,2021ApJ...913...27L,2022MNRAS.515.5071M,2023PhRvD.107d3005L,2023arXiv230508401M}.}.
This is the first study to examine the absolute stability of the 2SC quark star matter within NJL-type models, including the quark-antiquark scalar and vector interactions, as well as the diquark interaction. 

This paper is organized as follows. In Section~\ref{NJL model}, we introduce the formalism of the NJL-type model for describing the 2SC phase. 
Section~\ref{results} discusses the results on the dynamical quark mass and the color superconducting gap in 2SC quark matter, as well as the EOS with changing vector interaction and scalar diquark interaction. Our results are summarized in Section~\ref{summary}.

\section{The formalism}\label{NJL model}

\subsection{The two-flavor NJL-type model} \label{2f NJL model}

The SU(2) NJL-type model Lagrangian density, with the inclusion of quark-antiquark scalar and vector interactions and scalar diquark interaction, has the form as~\citep{1992RvMP...64..649K,2005PhR...407..205B}:
\begin{equation}
\begin{aligned}
\mathcal{L}= & \bar{q}\left(i \gamma^\mu \partial_\mu-m_0 +\mu \gamma^0\right) q+G_S\left[(\bar{q} q)^2+\left(\bar{q} i \gamma_5 \vec{\tau} q\right)^2\right] \\
- & G_V\left(\bar{q} \gamma^{\mu} q\right)^2 
+ G_D\left[\left( \bar{q} i \gamma_5 \tau_2 \lambda_A q_c\right)\left(
\bar{q}_c i \gamma_5 \tau_2 \lambda_A q\right)\right]\ . \label{NJL}
\end{aligned}
\end{equation}
Here, $q_c=C \bar{q}^{T},\  \bar{q}_c=q^T C$ are charge-conjugate spinors, $C=i \gamma^2 \gamma^0$ is the charge conjugation matrix (the superscript $T$ denotes the transposition operation), the quark field $q \equiv q_{i \alpha}$ with $i=1,2$ and $\alpha=1,2,3$ is a flavor doublet and color triplet, as well as a four-component Dirac spinor, $\lambda_A$ is asymmetric matrix in the color space. 
The effective coupling constants $G_S$, $G_V$, and $G_D$ are independent parameters that convey information about the strength of the scalar, vector interactions, and the diquark interaction near the Fermi surface, respectively.
$\mu$ is the quark chemical potential. When electric and color charge neutrality is considered, the chemical potential $\mu$ is a diagonal $6 \times 6$ matrix in flavor and color space, which can be expressed as $\mu=\operatorname{diag}\left(\mu_{ur}, \mu_{ug}, \mu_{ub}, \mu_{dr}, \mu_{dg}, \mu_{db}\right)$. $m_0$ is the diagonal mass matrix for quarks in flavor and color space, which contains
the small current quark masses and introduces a small explicit chiral symmetry breaking.

Following the Pauli principle, the diquark condensate demands the operator between two fermion fields in the diquark condensate to be asymmetric in Dirac, flavor, and color space. As a result, the diquark condensate can be represented as $\Delta \propto \varepsilon_{i j} \varepsilon^{\alpha \beta 3}$, where the Latin indices denote the flavors and the Greek indices represent the colors. The number ``3" signifies the chosen direction in color space.
We opt for blue as the preferred direction, thereby only the red and green colors participate in the condensate, while the blue one does not. Thus, the specific asymmetric operator is denoted as $\lambda_A=\lambda_2$.

Under the mean-field approximation, the Lagrangian can be written as:
\begin{equation}
\begin{aligned}
\widetilde{\mathcal{L}}&=  \bar{q}\left(i \gamma^\mu \partial_\mu- M +\tilde{\mu} \gamma^0\right) q \\
&-\frac{1}{2} \Delta^{*}\left(i \bar{q}_c \gamma_5 \tau_2 \lambda_2 q\right)
 -\frac{1}{2} \Delta \left(i \bar{q} \gamma_5 \tau_2 \lambda_2 q_c\right)\\
 &-\frac{(m_0 -M)^2}{4 G_S}+ \frac{(\mu-\tilde{\mu})^2}{4 G_V} -\frac{\Delta^{*} \Delta}{4 G_D} \ .  \label{MFA}
\end{aligned}
\end{equation}
Here, we introduce the constituent quark mass $M$ and effective quark chemical potential $\tilde{\mu}$ as follows:
\begin{equation}
\begin{aligned}
M & = m_0 - 2 G_S \sigma, \quad \sigma= \langle\bar{q} q\rangle \ , \\
\tilde{\mu} & =\mu-2 G_V n, \quad n= \left\langle\bar{q} \gamma^0 q\right\rangle \ , \\
\Delta & = -2 G_D \left\langle\bar{q}_c i \gamma_5 \tau_2 \lambda_2q\right\rangle \ , \\
\Delta^{*} & = -2 G_D \left\langle\bar{q} i \gamma_5 \tau_2 \lambda_2 q_c\right\rangle \ . \label{effective}
\end{aligned}
\end{equation}

\subsection{The thermodynamic potential}

In a star containing quark matter, it is necessary to ensure both beta equilibrium and charge neutrality.
The stellar matter is composed of quarks with the charge neutrality maintained by the inclusion of electrons.
Various quark chemical potentials are determined by their electric and color charges, given by $\mu_i = \mu - Q_e \mu_e + T_3 \mu_{3 c} + T_8 \mu_{8 c}$, with $Q_e$, $T_3$, and $T_8$ representing the generators of $U(1)_Q$, $U(1)_3$, and $U(1)_8$, respectively. 
Since the diquark condenses in the blue color direction and the other two colored quarks are degenerate, assuming $\mu_{3 c} = 0$ is safe~\citep{2003PhRvD..67f5015H,2007PhRvD..75b5024E}. 
The explicit expressions for the quark chemical potentials for each color and flavor are: 
\begin{equation}
\begin{aligned}
\mu_{ur} =\mu_{ug} =\frac{1}{3}\mu_B-\frac{2}{3} \mu_e+\frac{1}{3} \mu_{8 c} \ , \\
\mu_{dr} =\mu_{dg} =\frac{1}{3}\mu_B +\frac{1}{3} \mu_e+\frac{1}{3} \mu_{8 c} \ ,  \\
\mu_{ub} =\frac{1}{3}\mu_B-\frac{2}{3} \mu_e-\frac{2}{3} \mu_{8 c} \ ,  \\
\mu_{db} =\frac{1}{3}\mu_B+\frac{1}{3} \mu_e-\frac{2}{3} \mu_{8 c} \ . \label{Q chemical P}
\end{aligned}
\end{equation}

For quarks of the same flavor, the difference in chemical potentials between the red and green-colored quarks and the blue-colored quark is induced by $\mu_{8 c}$, while the difference in chemical potentials between $u$ and $d$ quarks, for the same color, is induced by $\mu_e$.
Hereafter for convenience, we define the mean chemical potential $\bar{\mu}$ for the pairing quarks, i.e., $q_{ur},\ q_{dg}$, and $q_{ug},\ q_{dr}$: 
\begin{equation}
\begin{aligned}
\bar{\mu}=\frac{\mu_{ur}+\mu_{dg}}{2}=\frac{\mu_{ug}+\mu_{dr}}{2}=\frac{1}{3}\mu_B -\frac{1}{6} \mu_e+\frac{1}{3} \mu_{8 c} \ , \label{mean potential}
\end{aligned}
\end{equation}
and the difference of the chemical potential $\delta \mu$ as: 
\begin{equation}
\begin{aligned}
\delta \mu=\frac{\mu_{dg}-\mu_{ur}}{2}=\frac{\mu_{dr}-\mu_{ug}}{2}=\frac{\mu_e}{2}\ . \label{difference Potential}
\end{aligned}
\end{equation}

With the help of the energy projectors for massive particles~\citep{2002PhRvD..65g6012H,2003PhRvD..67f5015H}, the thermodynamic potential of the 2SC quark-matter system can be derived as (see details in Appendix):
\begin{equation}
\begin{aligned}
\Omega_q &=  \frac{(m_0-M)^{2}}{4 G_S}-\frac{\left(\mu-\tilde{\mu}\right)^2}{4 G_V}+\frac{\Delta^2}{4 G_D}\\
&-2 \int \frac{d^3 \boldsymbol{p}}{(2 \pi)^3}\left\{2 E_{\boldsymbol{p}}+2 E_{\Delta}^{+}+2 E_{\Delta}^{-}\right. \\
& +T \ln \left[1+\exp \left(-\beta E_{ub}^{+}\right)\right]+T \ln \left[1+\exp \left(-\beta E_{ub}^{-}\right)\right] \\
& +T \ln \left[1+\exp \left(-\beta E_{db}^{+}\right)\right]+T \ln \left[1+\exp \left(-\beta E_{db}^{-}\right)\right] \\
& +2 T \ln \left[1+\exp \left(-\beta E_{\Delta^{+}}^{+}\right)\right]+2 T \ln \left[1+\exp \left(-\beta E_{\Delta^{-}}^{+}\right)\right] \\
& \left.+2 T \ln \left[1+\exp \left(-\beta E_{\Delta^{+}}^{-}\right)\right]+2 T \ln \left[1+\exp \left(-\beta E_{\Delta^{-}}^{-}\right)\right]\right\}
\end{aligned} \label{therm.}
\end{equation}
with $E_{\Delta^{\pm}}^{\pm}=E_{\Delta}^{\pm} \pm \delta \mu$, where $E_{\Delta}^{\pm} =\sqrt{(E_{\boldsymbol{p}} \pm \bar{\mu})^2 + \Delta^{2} } $, $E_{b}^{\pm}=E_{\boldsymbol{p}} \pm \tilde{\mu}_b$ and $E_{\boldsymbol{p}} =\sqrt{\boldsymbol{p}^2 +M^2}$. 

Assuming the electron mass is zero, one has the contribution from the electron gas  $\Omega_e$ as: 
\begin{equation}
\begin{aligned}
\Omega_e=-\frac{\mu_e^4}{12 \pi^2}\ . \label{electron gas}
\end{aligned}
\end{equation}
Then the total thermodynamic potential of the system is: 
\begin{equation}
\Omega=\Omega_q+\Omega_e\ . \label{total Thermo}
\end{equation}

\subsection{Gap equations and charge-neutrality conditions}

The self-consistent solutions of quark-antiquark condensate $\sigma$ and diquark condensate $\Delta$, as well as the quark number density $n$, correspond to the stationary points of the total thermodynamic potential: 
\begin{equation}
\begin{aligned}
\frac{\delta \Omega}{\delta M}=\frac{\delta \Omega}{\delta \Delta}=\frac{\delta \Omega}{\delta \tilde{\mu}}=\frac{\delta \Omega}{\delta \mu_{8c}}=\frac{\delta \Omega}{\delta \mu_{e}}=0 \ . \label{thermo condition}
\end{aligned}
\end{equation}
The gap equation for the constituent quark mass is obtained as:
\begin{equation}
\begin{aligned}
M= & m_0 +4 G_s M \int \frac{d^3 \boldsymbol{p}}{(2 \pi)^3} \frac{1}{E_{\boldsymbol{p}}}\left\{\left[1-f\left(E_{ub}^{+}\right)-f\left(E_{ub}^{-}\right)\right]\right.\\
&\left.+\left[1-f\left(E_{db}^{+}\right)-f\left(E_{db}^{-}\right)\right]\right. \\
& \left.+2 \frac{E_{\boldsymbol{p}}^{+}}{E_{\Delta}^{+}}\left[1-f\left(E_{\Delta^{+}}^{+}\right)-f\left(E_{\Delta^{-}}^{+}\right)\right]\right.\\
&\left.+2 \frac{E_{\boldsymbol{p}}^{-}}{E_{\Delta}^{-}}\left[1-f\left(E_{\Delta^{+}}^{-}\right)-f\left(E_{\Delta^{-}}^{-}\right)\right]\right\}\ . \label{Mass gap}
\end{aligned}
\end{equation}
where $E_{\boldsymbol{p}}^{\pm}= E_{\boldsymbol{p}} \pm \bar{\mu}$ and $\tilde{f}(x)=1 /\left(e^{\beta x}+1\right)$ is the Fermi-Dirac distribution function with $\beta=1/T$. 

The obtained gap equation for diquark condensate is: 
\begin{equation}
\begin{aligned}
\Delta= & 4 G_D \Delta \int \frac{d^3 \boldsymbol{p}}{(2 \pi)^3} \left\{ \frac{2}{E_{\Delta}^{-}}\left[1-f\left(E_{\Delta^{+}}^{-}\right)-f\left(E_{\Delta^{-}}^{-}\right)\right]\right. \\
& \left.+ \frac{2}{E_{\Delta}^{+}}\left[1-f\left(E_{\Delta^{+}}^{+}\right)-f\left(E_{\Delta^{-}}^{+}\right)\right]\right\}\ .\label{diquark gap}
\end{aligned}
\end{equation}
By utilizing the thermodynamic potential in terms of $\tilde{\mu}$ in Eq.~(\ref{thermo condition}), we can obtain the number density of the two colored up quarks,
\begin{equation}
\begin{aligned}
n_{ur}=n_{ug}= & \int \frac{d^3 \boldsymbol{p}}{(2 \pi)^3} \left\{\frac{E^{+}}{E_{\Delta}^{+}}\left[1-f\left(E_{\Delta^{+}}^{+}\right)-f\left(E_{\Delta^{-}}^{+}\right)\right]\right. \\
- & \left.\frac{E^{-}}{E_{\Delta}^{-}}\left[1-f\left(E_{\Delta^{+}}^{-}\right)-f\left(E_{\Delta^{-}}^{-}\right)\right] \right.\\
+ &\left.f\left(E_{\Delta^{+}}^{-}\right)+f\left(E_{\Delta^{+}}^{+}\right)-f\left(E_{\Delta^{-}}^{+}\right)-f\left(E_{\Delta^{-}}^{-}\right)\right\} \ ,\label{u quark N densities}
\end{aligned}
\end{equation}
and the two colored down quarks, 
\begin{equation}
\begin{aligned}
n_{dr}=n_{dg}= & \int \frac{d^3 \boldsymbol{p}}{(2 \pi)^3} \left\{\frac{E^{+}}{E_{\Delta}^{+}}\left[1-f\left(E_{\Delta^{+}}^{+}\right)-f\left(E_{\Delta^{-}}^{+}\right)\right]\right. \\
-& \left.\frac{E^{-}}{E_{\Delta}^{-}}\left[1-f\left(E_{\Delta^{+}}^{-}\right)-f\left(E_{\Delta^{-}}^{-}\right)\right]\right. \\
-&  \left.f\left(E_{\Delta^{+}}^{-}\right)-f\left(E_{\Delta^{+}}^{+}\right)+f\left(E_{\Delta^{-}}^{+}\right)+f\left(E_{\Delta^{-}}^{-}\right)\right\}\ .\label{d quark N densities}
\end{aligned}
\end{equation}
For the unpaired blue quarks, we have:
\begin{equation}
\begin{aligned}
& n_{ub}=2 \int \frac{d^3 \boldsymbol{p}}{(2 \pi)^3} \left[f\left(E_{ub}^{-}\right)-f\left(E_{ub}^{+}\right)\right]\ , \\
& n_{db}=2 \int \frac{d^3 \boldsymbol{p}}{(2 \pi)^3} \left[f\left(E_{db}^{-}\right)-f\left(E_{db}^{+}\right)\right]\ .
\end{aligned}
\end{equation}
The color charge-neutrality condition requires the system to have zero net charge $T_8$.
This requirement is met by evaluating the equation $n_8=-\frac{\delta \Omega}{\partial \mu_{8 c}}=0$, resulting in the color charge-neutrality condition:
\begin{equation}
\begin{aligned}
& \int \frac{d^3 \boldsymbol{p}}{(2 \pi)^3}\left\{ \frac{E^{+}}{E_{\Delta}^{+}}\left[1-f\left(E_{\Delta^{+}}^{+}\right)-f\left(E_{\Delta^{-}}^{+}\right)\right]\right. \\
& - \frac{E^{-}}{E_{\Delta}^{-}}\left[1-f\left(E_{\Delta^{+}}^{-}\right)-f\left(E_{\Delta^{-}}^{-}\right)\right] \\
& \left.+f\left(E_{ub}^{+}\right)-f\left(E_{ub}^{-}\right)+f\left(E_{db}^{+}\right)-f\left(E_{db}^{-}\right)\right\}=0 \ . \label{color neutrality}
\end{aligned}
\end{equation}
The electric chemical potential $\mu_e$ 
should be chosen to ensure the condition of electric charge neutrality. From $n_Q=-\frac{\delta \Omega}{\delta \mu_e}=0$, one has:
\begin{equation}
\begin{aligned}
& \int \frac{d^3 \boldsymbol{p}}{(2 \pi)^3}\left\{2 \frac{E^{+}}{E_{\Delta}^{+}}\left[1-f\left(E_{\Delta^{+}}^{+}\right)-f\left(E_{\Delta^{-}}^{+}\right)\right]\right. \\
& \left.-2 \frac{E^{-}}{E_{\Delta}^{-}}\left[1-f\left(E_{\Delta^{+}}^{-}\right)-f\left(E_{\Delta^{-}}^{-}\right)\right] \right.\\
&\left.+6\left[ f\left( E_{\Delta^{+}}^{-}\right) +f\left( E_{\Delta^{+}}^{+}\right) -f\left( E_{\Delta^{-}}^{+}\right) -f\left( E_{\Delta^{-}}^{-}\right)    \right]  \right.\\
& \left.+4\left[f\left(E_{ub}^{-}\right)-f\left(E_{ub}^{+}\right)\right]-2\left[f\left(E_{db}^{-}\right)-f\left(E_{db}^{+}\right)\right]\right\}-\frac{\mu_e^3}{\pi^2}=0 \ .\label{electric neutrality}
\end{aligned}
\end{equation}
\begin{figure}
\centering
\includegraphics[width=0.49\textwidth]{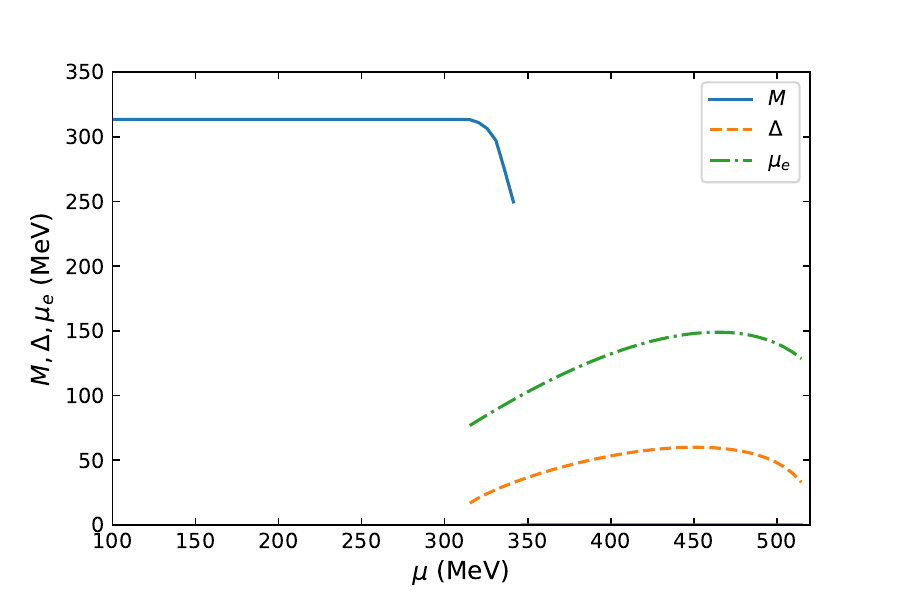}
    \vskip-2mm
\caption{
The constituent quark mass $M$, the electric chemical potential $\mu_e$ and the diquark gap $\Delta$ as functions of the quark chemical potential for $R_V=0.0$, $R_D=0.75$.}\label{fig:Mmu0.75}
    \vspace{-0.4cm}
\end{figure}
\begin{figure}
\centering
\includegraphics[width=0.49\textwidth]{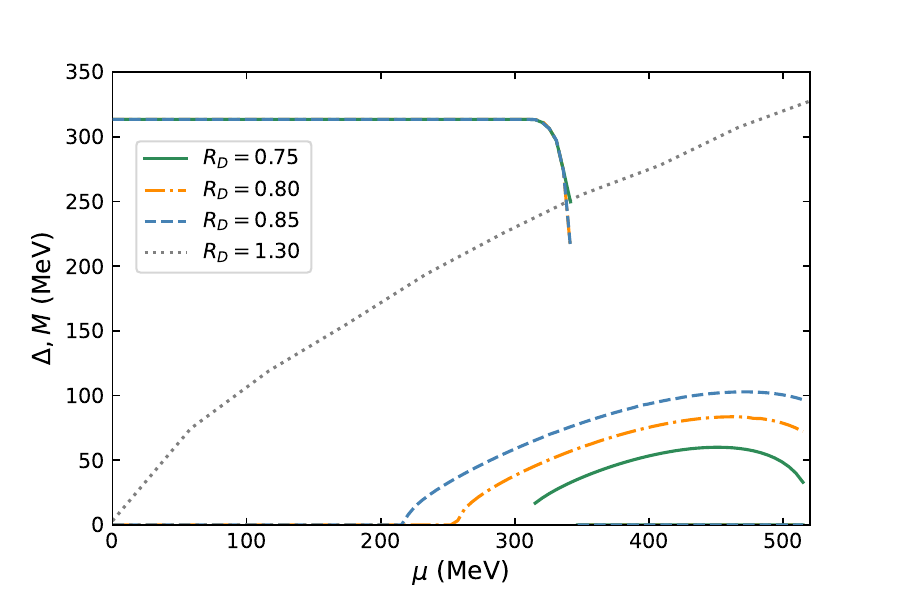}
    \vskip-2mm
\caption{The constituent quark mass $M$ and the diquark gap $\Delta$ as functions of the quark chemical potential at different scalar diquark coupling constants $R_D=0.75,0.80,0.85,1.30$ for fixed vector interaction $R_V=0$.
}\label{fig:Mmu}
    \vspace{-0.4cm}
\end{figure}

\subsection{Parameter fixing in NJL model}

Before proceeding with calculations, parameter fitting is necessary.
Since the NJL model is not renormalizable, a momentum cutoff $\Lambda$ needs to be introduced. In the chiral limit ($m_0=0$), the fermion momentum cutoff $\Lambda=0.6533\gev$ and the scalar quark-antiquark coupling constant $G_S=5.0163 \gev^{-2}$ in the color singlet channel are determined by fitting experimental data on the pion decay constant and pion mass. 
The coupling strengths of $G_V$ and $G_D$ cannot currently be fixed and involve many uncertainties.
The value of $G_V/G_S$ ranges from $0.25$ to $0.5$, derived by a Fierz transformation of the effective one-gluon exchange interaction, where $G_{V}$ depends on the strength of the $U_{A}(1)$ anomaly in the two-flavor model~\citep{1992RvMP...64..649K,2011PhRvD..84e6010K}. 
Some studies suggest that $G_{V}$ is similar in magnitude to $G_D$ to explain lattice results on the curvature of the linear chiral restoration at zero density~\citep{2005PhR...407..205B,2013PhLB..719..131B,2013ApJ...764...12M}.
The strength of $G_D$ is even less clear. 
Typically, one might assume $G_D/G_S=3/4$ for equal contributions from the quark-antiquark interaction channels and the Fierz-transformed diquark interaction channels~\citep{2005PhRvD..72e4024S,2014RvMP...86..509A,2015PhRvD..92j5030C}.
In the present study, we treat both $G_V$ and $G_D$ as free parameters, for our purpose of exploring whether a parameter space exists for absolutely stable 2SC quark matter.
For later convenience, we define $G_V/G_S\equiv R_V$ and $G_D/G_S\equiv R_D$. 

\section{Results and discussion}
\label{results}

\subsection{Dynamically generated mass and pairing gaps}

In Fig.~\ref{fig:Mmu0.75}, we present the constituent quark mass and pairing gap
for 2SC quark matter under electric and color charge neutrality conditions, considering a typical case with $R_V=0$, $R_D=0.75$. 
It is shown that when the quark chemical potential exceeds the vacuum constituent quark mass ($M=313\mev$), chiral symmetry begins to restore. 
The 2SC phase emerges around $\mu=313\mev$, in which diquark condensate $\Delta$ 
subsequently decreases after reaching its maximum. 
This cessation in the increasing trend is related to the momentum cutoff.
In this charged 2SC phase, where the electrons have a large Fermi surface 
extending to $\sim150\mev$, the magnitude of $\Delta$ is around $50\mev$, 
significantly smaller than the uncharged case ($\sim100\mev$)~\citep{2002PhRvD..65g6012H,2003PhRvD..67f5015H,2023PhRvD.108d3008Y}. 
The distinct Fermi surfaces of the two pairing quarks contribute to the reduction in the magnitude of the diquark condensate.

In Fig.~\ref{fig:Mmu}, we specifically examine the influence of the strength of the attractive diquark interaction.
It is seen that, when the vector interaction is absent ($R_V=0$), an increase in the attractive diquark interaction strength leads to a larger maximum value of the diquark gap ($\Delta$), while the critical chemical potential, marking the onset of the 2SC phase, decreases.
The nature of the chiral phase transition remains first order, and its transition hardly changes.
However, for extremely large $R_D$ coupling (as large as e.g., $R_D=1.3$), 
the 2SC phase emerges at significantly lower quark chemical potentials, approaching zero; Simultaneously, the maximum $\Delta$ value exceeds the constituent quark mass ($M$), rendering it physically unrealistic.

\subsection{The EOS}

At finite chemical potential and zero temperature, the pressure-chemical potential relation for quark matter can be rigorously derived using the functional path integrals of QCD~\citep{2008PhRvD..78e4001Z}: 
\begin{equation}
P(\mu ; M)=P(\mu=0 ; M)+\int_0^\mu d \mu^{\prime} \rho\left(\mu^{\prime}\right)\ . \label{P mu}
\end{equation}
While the negative vacuum pressure $P(\mu=0 ; M)$ can be consistently calculated in NJL-type models~\citep{2005PhR...407..205B}, the results are unsatisfactory due to the lack of confinement at vanishing density, as discussed in detail in~\citet{2022PhRvD.105l3004Y}. 
Therefore, in this study, we treat $P(\mu=0 ; M)$ as a phenomenological parameter corresponding to $-B$ (with $B$ being the vacuum bag constant), which preserves quark confinement.

\begin{figure}
\centering
\includegraphics[width=0.49\textwidth]{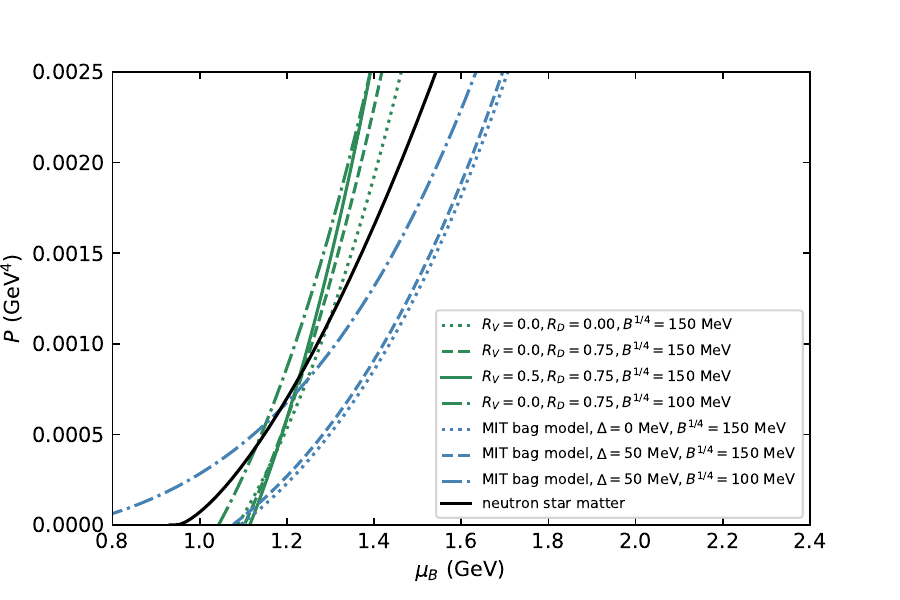}
    \vskip-2mm
\caption{Pressure as a function of the baryon chemical potential for normal quark matter and 2SC quark matter. The green curves are the corresponding results from the present NJL-type model, while the blue curves represent the MIT bag model calculations for $a_4=0.61$~\citep{2002JHEP...06..031A,2021PhRvD.103f3018Z}. A representative result for neutron star matter within quark mean-field model~\citep{2020JHEAp..28...19L} is also shown with a solid black curve.
}\label{fig:PmuB}
    \vspace{-0.4cm}
\end{figure} 

The thermodynamic relation between energy density and pressure is the EOS of the system
\begin{equation}
\varepsilon=-P(\mu ; M)+\sum_{i=u, d, s, e} \mu_i \rho_i\left(\mu_i\right)\ .
\end{equation}
The pressure of normal quark matter and 2SC quark matter within NJL-type models are shown together in Fig.~\ref{fig:PmuB}.
In the same plot, we include the corresponding results from MIT bag model calculations~\citep{2002JHEP...06..031A,2021PhRvD.103f3018Z}, where we have purposely chosen a similar diquark gap ($\Delta=50 \mev$; see discussions above accompanying Figs.~\ref{fig:Mmu} and \ref{fig:Mmu0.75}).
Additionally, we include a representative result for neutron star matter within the quark mean-field model~\citep{2020JHEAp..28...19L}. 

It is seen that in both MIT and NJL-type models, 
with increasing chemical potential, the pressure of 2SC quark matter becomes higher than that of normal quark matter. This indicates the lower energy of 2SC matter, making it energetically favored. 
Specially in NJL-type model calculations, the inclusion of vector interaction pushes the baryon chemical potential of quark matter at zero pressure to high baryon densities, resembling (although not as significant as) an increase in $-B$ (a decrease in vacuum bag constant). 
Conversely, the inclusion of diquark interaction has the opposite effect in both MIT and NJL-type models, due to its attractive nature.
The results from MIT model calculations generally appear energetically unfavorable, staying below those of NJL-type model calculations and even below those of neutron star matter. In contrast, for the NJL-type model, starting from a certain value of baryon chemical potential ($\mu_B \sim 1.1\gev$), the pressure of neutron star matter is lower than that of quark matter. 
This lends initial support to the possibility of the existence of self-bound compact stars composed of deconfined quarks, as mentioned in our introduction.

\subsection{Is the compact object associated with HESS J1731-347 a 2SC quark star?}

\begin{figure}
\centering
\includegraphics[width=0.49\textwidth]{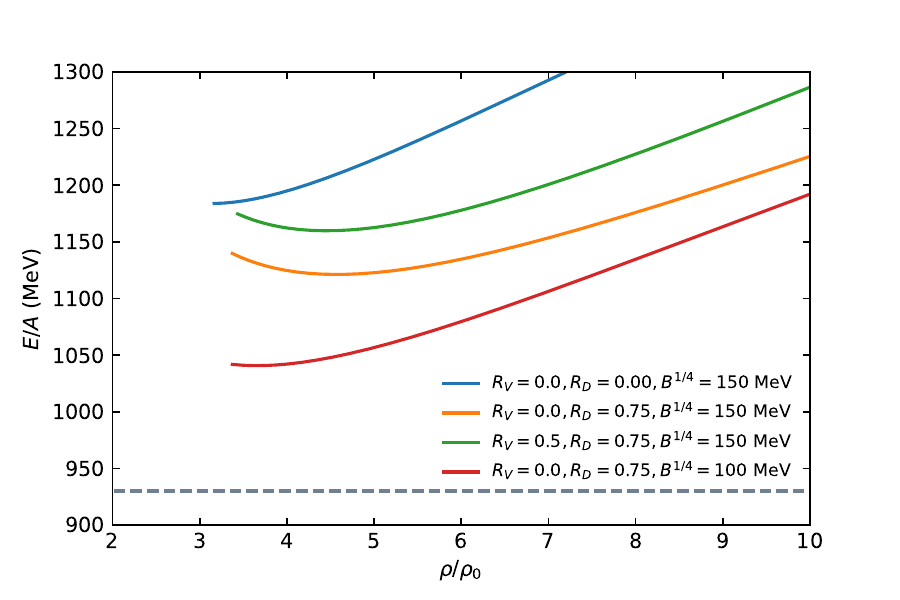}
    \vskip-2mm
\caption{Energy per baryon versus baryon density (in units of the nuclear saturation density $\rho_0$) 
for select sets of parameters in NJL-type models. The horizontal line represents the energy per baryon of the most stable nuclei known, $E/A (^{56}\rm Fe)=930\mev$. 
}\label{fig:EnerPerBaryon} 
    \vspace{-0.4cm}
\end{figure}
\begin{figure}
\centering
\includegraphics[width=0.49\textwidth]{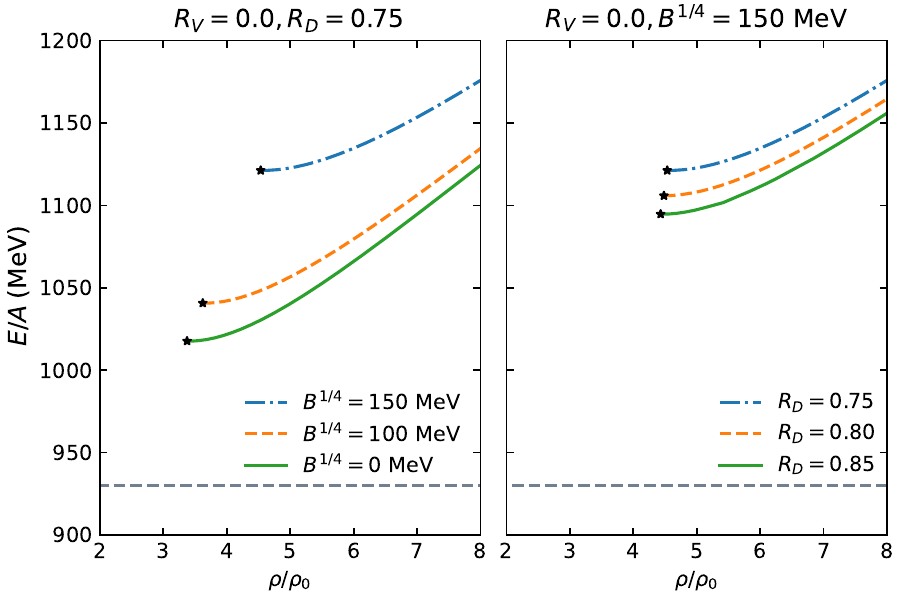}
    \vskip-2mm
\caption{Left penal: Effect of vacuum bag constant $B$ on energy per baryon with $R_V=0$, $R_D=0.75$ at $B^{1/4}=0, 100, 150\mev$. Right penal: Effect of the strength of scalar diquark interaction on energy per baryon with $R_V=0$, $B^{1/4}=150\mev$ at $R_D=0.75, 0.80, 0.85$. The starting star point of each line indicates the stellar surface at zero pressure. 
}\label{fig:EAB} 
    \vspace{-0.4cm}
\end{figure}
\begin{figure}
\centering
\includegraphics[width=0.49\textwidth]{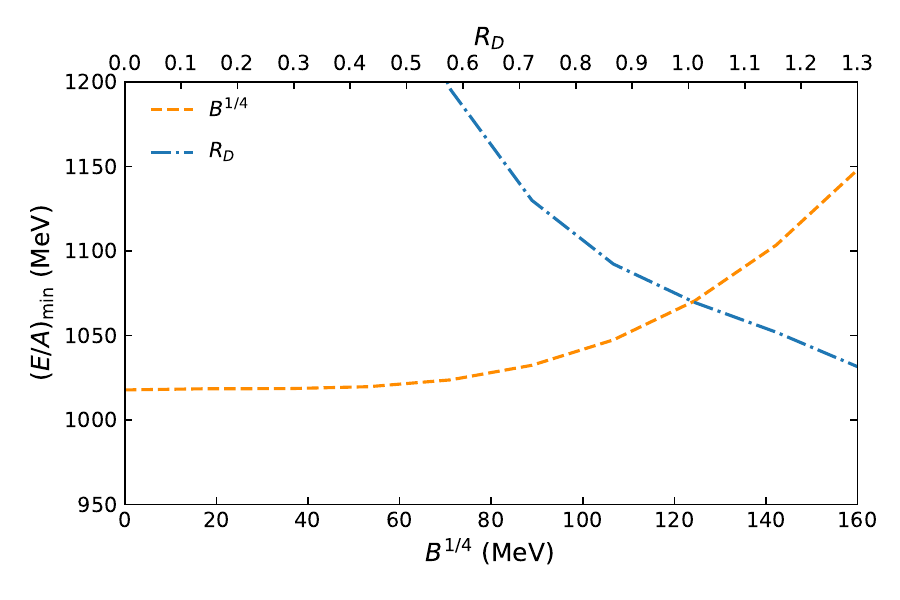}
    \vskip-2mm
\caption{Minimum energy per baryon $(E/A)_{\text{min}}$ as functions of the vacuum bag constant $B^{1/4}$ (orange dashed curve) and diquark interacting strength $R_D$ (blue dash-dotted curve).
}\label{fig:EAminGDB}
    \vspace{-0.4cm}
\end{figure}
Recently, the reported presence of a very low-mass compact star in the supernova remnant HESS J1731-347 has stimulated many investigations in the light of quark stars. 
Quark stars exist when one applies the Bodmer-Witten hypothesis~\citep{1971PhRvD...4.1601B,1984PhRvD..30..272W}. 
See discussions also in ~\citet{2018PhRvL.120v2001H,2018PhRvD..97h3015Z,2020PhRvD.102h3003R,2020PhRvD.101d3003Z,2021MNRAS.506.5916L,2021ApJ...917L..22M,2022PhRvD.105l3004Y}.
Correspondingly, the NJL-type model parameters should be chosen to ensure that the energy per baryon ($E/A$) of 2SC quark matter must be lower than $930\mev$, i.e., the $E/A$ of the most stable iron nuclei.

Fig.~\ref{fig:EnerPerBaryon} shows the $E/A$ of quark matter under various NJL model parameters. 
We see again the opposite effects of increasing the bag constant, vector interaction, and diquark interaction on the stability of quark matter.Fig.~\ref{fig:EAB} delve deeper into the effects of the vacuum bag parameter $B$ (in the left panel) and gap parameter $\Delta$ of color superconductor (in the right panel), in lowering the energy per baryon of quark matter and decreasing quark star surface density at zero pressure.
Specifically, we highlight the minimum energy per baryon density $(E/A)_{\rm{min}}$ as functions of $B^{1/4}$ and $R_D$ in Fig.~\ref{fig:EAminGDB}.
$(E/A)_{\rm{min}}$ of 2SC quark matter exhibits higher sensitivity to larger vacuum pressure. However, for smaller values, such as $B^{1/4}$ with $B^{1/4} \lesssim 60\mev$, the minimum energy per baryon remains relatively constant. 
While increasing the attractive interactions $R_D$ continues to lower the energy per baryon, excessively large diquark interactions result in unrealistic gap parameters, as demonstrated in Fig.~\ref{fig:Mmu}. Consequently, there is no viable parameter for $B^{1/4}$ that can render 2SC quark matter more stable than the most stable iron nuclei. In other words, 2SC quark matter cannot be considered absolutely stable.
Note that previously, \citet{2002JHEP...06..031A} demonstrated the absence of a 2SC 
 phase within the bag model in the interior of compact stars in three-flavor quark matter with the inclusion of unpaired strange quarks. 

\section{Summary}\label{summary}

In this study, we investigated the possibility of absolutely stable 2SC quark matter within NJL-type models, adopting the three-momentum cutoff regularization scheme while incorporating vector and scalar diquark interaction. 
The study shed light on the substantial impact of repulsive vector interactions and attractive diquark interactions near the Fermi surface on the stability of 2SC quark matter within the NJL-type model.

The results indicated that the introduction of charge neutrality in the 2SC phase suppresses the size of the superconducting energy gap $\Delta$ compared to the phase without charge neutrality. 
As the 2SC phase emerges at lower chemical potentials, the maximum value of the color superconducting gap increases. 
While an increase in diquark interaction minimally affects the chiral phase transition, it significantly influences the critical quark chemical potential at which the 2SC phase initiates.
Despite the lowering of vacuum pressure $B^{1/4}$ and the diquark interaction $R_D$ resulting in a decreased $E/A$, no viable parameter space exists for absolutely stable 2SC quark matter, specifically where $E/A~(\rm{2SC}) < 930\mev~(^{56}\rm Fe)$.
Consequently, the presence of 2SC quark matter in the compact object associated with HESS J1731-347 is not supported by the present stability study within NJL-type models.
For further extension of our work, it may be interesting to incorporate the effects of gluons~\citep{2000PhRvD..62c4007R,2000NuPhB.582..571C,2005NuPhB.723..255G,2015PhRvD..92d3010F} and finite temperature~\citep{2000PhRvD..61e1501P,2003NuPhA.729..835H,2007PhRvD..75c6003H}.

We mention here that there are several alternative models discussed in the literature for HESS J1731-347, such as (hybrid) neutron stars~\citep{2023PhRvD.107b3012T,2023PhLB..84438062L,2023PhRvC.108d5803K,2023PhRvC.108b5806B,2023arXiv230604992H,2023arXiv231207113L}, and dark-matter-admixed neutron stars~\citep{2023arXiv230712748R,2023ApJ...958...49S}.
In practice, measuring the mass and radius of compact stars is an extremely challenging task~\citep[see more discussions in e.g.,][]{2023PhRvD.108i4014B,2023ApJ...949...11J,miao24}. We eagerly anticipate more updated results in the near future, especially as models for atmosphere, distance, and other factors continue to develop.

\appendix
The appendix shows in detail how to evaluate the thermodynamic potential of the 2SC quark-matter system. 
\\
\section{Partition function and the thermodynamic potential}

The partition function of the grand canonical ensemble can be evaluated by using the standard method:
\begin{equation}
\begin{aligned}
\mathcal{Z}=N^{\prime} \int[d \bar{q}][d q] \exp \left\{\int_0^\beta d \tau \int d^3 \boldsymbol{x}\left(\mathcal{L}+\mu \bar{q} \gamma_0 q\right)\right\}\ .
\end{aligned}
\end{equation}
Under mean-field approximation, the Lagrangian in Eq.~(\ref{NJL}) becomes Eq.~(\ref{MFA}). 
After introducing the bispinor field $\Psi(x)$, the effective Lagrangian can be written in the momentum space as follows:
\begin{equation}
\begin{aligned}
\widetilde{\mathcal{L}}=\bar{\Psi} G^{-1} \Psi+V \ .
\end{aligned}
\end{equation}
Here, the inverse of the propagator matrix is:
\begin{equation}
\begin{aligned}
G^{-1}=\left(\begin{array}{cc}
\gamma^{\mu}p_{\mu}+\tilde{\mu} \gamma^0-M & \Delta^- \\
 \Delta^+& \gamma^{\mu}p_{\mu}-\tilde{\mu} \gamma^0-M
\end{array}\right) \label{eq quark propagator Appdenix}
\end{aligned}
\end{equation}
with $\Delta^-=-i\Delta \gamma_5 \tau_2 \lambda_2 $ and $\Delta^+=-i\Delta^{*} \gamma_5 \tau_2 \lambda_2$.
The quark propagator in the normal phase is:
\begin{equation}
\begin{aligned}
\mathrm{G}_0^{-1}=\left(\begin{array}{cc}
{\left[G_0^{+}\right]^{-1}} & 0 \\
0 & {\left[G_0^{-}\right]^{-1}}
\end{array}\right)\ ,
\end{aligned} \label{normal}
\end{equation}
with $\left[G_0^{\pm}\right]^{-1}=\left(p_0 \pm \tilde{\mu} \right) \gamma_0-\boldsymbol{\gamma} \cdot \boldsymbol{p}-M$
and $p_0=i \omega_n,\ \omega_n=(2n +1)\pi T$.
The quark propagator in the color-breaking phase is:
\begin{equation}
\begin{aligned}
\mathrm{G}^{-1}=\left(\begin{array}{cc}
{\left[G_0^{+}\right]^{-1}} & \Delta^{-} \\
\Delta^{+} & {\left[G_0^{-}\right]^{-1}}
\end{array}\right) \ .
\end{aligned}
\end{equation}
By solving $\mathrm{G}^{-1} \mathrm{G}=1$, the Nambu-Gorkov propagator $\mathrm{G}(p)$ can be determined, resulting in
\begin{equation}
\begin{aligned}
\mathrm{G}=\left(\begin{array}{ll}
G^{+} & \Xi^{-} \\
\Xi^{+} & G^{-}
\end{array}\right) \ ,
\end{aligned}
\end{equation}
with the components
\begin{equation}
\begin{aligned}
& G^{\pm} \equiv\left\{\left[G_0^{\pm}\right]^{-1}-\Sigma^{\pm}\right\}^{-1}, \quad \Sigma^{\pm} \equiv \Delta^{\mp} G_0^{\mp} \Delta^{\pm},  \quad 
 \Xi^{\pm} \equiv-G^{\mp} \Delta^{\pm} G_0^{\pm}=-G_0^{\mp} \Delta^{\pm} G^{\pm} \ ,
\end{aligned}
\end{equation}
in which all components depend on the 4-momentum $p^\mu$.

As we mentioned in Sec.~\ref{2f NJL model}, because the blue quarks of $u$ and $d$ do not participate in the diquark condensate, the partition function can be written as a product of the following three parts:
\begin{equation}
\begin{aligned}
\mathcal{Z}=\mathcal{Z}_{\text {const }} \mathcal{Z}_{b} \mathcal{Z}_{r,g} \ , \label{partition function}
\end{aligned}
\end{equation}
where the constant part is:
\begin{equation}
\begin{aligned}
\mathcal{Z}_{\text {const }}=N^{\prime} \exp \left\{-\int_0^\beta d \tau \int d^3 \boldsymbol{x}\left[\frac{\sigma^2}{4 G_S}-\frac{(\mu-\tilde{\mu})^2}{4 G_V}+\frac{\Delta^* \Delta}{4 G_D}\right]\right\} \ .
\end{aligned}
\end{equation}
In Eq.~(\ref{partition function}), the $\mathcal{Z}_{b}$ part is for the unpaired quarks $u_b$ and $d_b$, and the $\mathcal{Z}_{r,g}$ part is for the quarks participating in pairing, i.e., red $u$ quark ($u_r$) paired with green $d$ quark ($d_g$), and green $u$ quark ($u_g$) paired with red $d$ quark ($d_r$). 
In the following, we will derive the contribution of $\mathcal{Z}_{b}$ and $\mathcal{Z}_{r,g}$.

\subsection{The energy projectors for massive particles and dispersion relations}

To handle the intricate massive quark propagator, we can employ energy projectors for massive particles to evaluate it.
The energy projectors onto positive- and negative-energy states for free massive particles are defined as follows:
\begin{equation}
\begin{aligned}
\Lambda_{\pm}(\boldsymbol{p})=\frac{1}{2}\left(1 \pm \frac{\gamma_0(\boldsymbol{\gamma} \cdot \boldsymbol{p}+M)}{E_{\boldsymbol{p}}}\right)\ ,
\end{aligned}
\end{equation}
in which the quark energy is $E_{\boldsymbol{p}}=\sqrt{\boldsymbol{p}^2+M^2}$. Under the transformation of $\gamma_0$ and $\gamma_5$, we can get another two energy projectors
\begin{equation}
\begin{aligned}
\tilde{\Lambda}_{\pm}(\boldsymbol{p})=\frac{1}{2}\left(1 \pm \frac{\gamma_0(\boldsymbol{\gamma} \cdot \boldsymbol{p}-M)}{E_{\boldsymbol{p}}}\right)\ ,
\end{aligned}
\end{equation}
which satisfy
\begin{equation}
\begin{aligned}
\gamma_0 \Lambda_{\pm} \gamma_0=\tilde{\Lambda}_{\mp}, \quad \gamma_5 \Lambda_{\pm} \gamma_5=\tilde{\Lambda}_{\pm} \ .
\end{aligned}
\end{equation}
Using energy projectors, the quark propagator for the normal phase [Eq.~(\ref{normal})] can be rewritten as:
\begin{equation}
\begin{aligned}
G_0^{\pm} & =\frac{1}{\gamma_0 (p_0 \pm \tilde{\mu} )-\boldsymbol{\gamma} \cdot \boldsymbol{p} -M}\\
&=\frac{\gamma_0 (p_0 \pm \tilde{\mu})-\boldsymbol{\gamma} \cdot \boldsymbol{p} + M}{[\gamma_0 (p_0 \pm \tilde{\mu})-\boldsymbol{\gamma} \cdot \boldsymbol{p} -M] [\gamma_0 (p_0 \pm \tilde{\mu})-\boldsymbol{\gamma} \cdot \boldsymbol{p} + M]} \\
&=\frac{\gamma_0\left(p_0 \pm \tilde{\mu} \right)-\boldsymbol{\gamma} \cdot \boldsymbol{p}+ M}{\left(p_0+E_{\boldsymbol{p}}^{\pm}\right)\left(p_0-E_{\boldsymbol{p}}^{\mp}\right)} 
 =\frac{\gamma_0\left(p_0 \pm \tilde{\mu} \right)-\gamma_0\left(E_{\boldsymbol{p}} \tilde{\Lambda}_{+}-E_{\boldsymbol{p}} \tilde{\Lambda}_{-}\right)}{\left(p_0+E_p^{\pm}\right)\left(p_0-E_{\boldsymbol{p}}^{\mp}\right)} 
 =\frac{\gamma_0 \tilde{\Lambda}_{+}}{p_0+E_{\boldsymbol{p}}^{\pm}}+\frac{\gamma_0 \tilde{\Lambda}_{-}}{p_0-E_{\boldsymbol{p}}^{\mp}} \ ,
\end{aligned}
\end{equation}
with $E_{\boldsymbol{p}}^{\pm}=E_{\boldsymbol{p}} \pm \tilde{\mu}$.
Then, the quark propagator in Eq.~(\ref{eq quark propagator Appdenix}) which includes the contribution of diquark condensate can be evaluated as: 
\begin{equation}
\begin{aligned}
{\left[G^{\pm}\right]^{-1} } & =\left[G_0^{+}\right]^{-1}-\Sigma \\
& =\gamma_0\left[p_0-E_{\boldsymbol{p}}^{\mp}\right] \Lambda_{+}+\gamma_0\left[p_0+E_{\boldsymbol{p}}^{\pm}\right] \Lambda_{-}
-\frac{\Delta^{-} \gamma_0 \tilde{\Lambda}_{+} \Delta^{+}}{p_0+E_{\boldsymbol{p}}^{\mp}}
+\frac{\Delta^{-} \gamma_0 \tilde{\Lambda}_{-} \Delta^{+}}{p_0-E_{\boldsymbol{p}}^{\pm}} \\
& =\frac{\left[p_0-E_{\boldsymbol{p}}^{\pm}\right]\left\{\left[p_0^2-E_{\boldsymbol{p}}^{\mp^2}-\Delta^2\right] \gamma_0 \Lambda_{+}\right\}}{\left(p_0-E_{\boldsymbol{p}}^{\pm}\right)\left(p_0+E_{\boldsymbol{p}}^{\mp}\right)}
+\frac{\left[p_0+E_{\boldsymbol{p}}^{\mp}\right]\left\{\left[p_0^2-E_{\boldsymbol{p}}^{\pm^2}-\Delta^2\right] \gamma_0 \Lambda_{-}\right\}}{\left(p_0-E_{\boldsymbol{p}}^{\pm}\right)\left(p_0+E_{\boldsymbol{p}}^{\mp}\right)} \ .
\end{aligned}
\end{equation}
By solving $G^{-1}G=1$, we can determine the Nambu-Gorkov propagator
\begin{equation}
\begin{aligned}
G^{\pm}=
\frac{\left(p_0-E_{\boldsymbol{p}}^{\pm}\right)\left(p_0+E_{}\boldsymbol{p}^{\mp}\right)}{\left[p_0-E_{\boldsymbol{p}}^{\pm}\right]\left\{\left[p_0^2-E_{\Delta}^{\mp}\right] \gamma_0 \Lambda_{+}\right\}+\left[p_0+E_{\boldsymbol{p}}^{\mp}\right]\left\{\left[p_0^2-E_{\Delta}^{\pm^2}\right] \gamma_0 \Lambda_{-}\right\}} \ .
\end{aligned}
\end{equation}
%
Multiplying both the dominator and numerator of the above equations by
\begin{equation}
\begin{aligned}
\left(p_0^2-E_{\Delta}^{\pm 2}\right)\left(p_0+E_{\boldsymbol{p}}^{\mp}\right) \gamma_0 \tilde{\Lambda}_{-}+\left(p_0^2-E_{\Delta}^{\mp 2}\right)\left(p_0-E_{\boldsymbol{p}}^{\pm}\right) \gamma_0 \tilde{\Lambda}_{+} \ ,
\end{aligned}
\end{equation}
we can obtain the following expression:
\begin{equation}
\begin{aligned}
G^{\pm}=\frac{\left(p_0+E_{\boldsymbol{p}}^{\mp}\right) \gamma_0 \tilde{\Lambda}_{-}}{p_0^2-E_{\Delta}^{\mp}}+\frac{\left(p_0-E_{\boldsymbol{p}}^{\pm}\right) \gamma_0 \tilde{\Lambda}_{-}}{p_0^2-E_{\Delta}^{\pm^2}} \ .
\end{aligned}
\end{equation}
Doing a similar calculation, we get
\begin{equation}
\begin{aligned}
\Xi^{\pm}=\frac{\Delta^{\pm} \tilde{\Lambda}_{-}}{p_0^2-E_{\Delta}^{\mp^2}}+\frac{\Delta^{\pm} \tilde{\Lambda}_{-}}{p_0^2-E_{\Delta}^{\pm^2}} \ ,
\end{aligned}
\end{equation}
with $E_{\Delta}^{\pm 2}=E_{\boldsymbol{p}}^{\pm 2}+\Delta^2$. 

\subsection{Calculating the blue quarks' partition function}

For the two blue quarks not participating in the diquark condensate, we have
\begin{equation}
\begin{aligned}
\mathcal{Z}_{b} & =\int\left[d \Psi_{b}\right] \exp \left\{\frac{1}{2} \sum_{n, p} \bar{\Psi}_{b} \frac{\left[G_0^{-1}\right]_{b}}{T} \Psi_{b}\right\} 
 =\operatorname{Det}^{1 / 2}\left(\beta\left[G_0^{-1}\right]_{b}\right)\ ,
\end{aligned}
\end{equation}
and
\begin{equation}
\begin{aligned}
\ln \mathcal{Z}_{b}= & \frac{1}{2} \ln \operatorname{Det}\left(\beta\left[G_0^{-1}\right]_{b}\right) \\
= & \frac{1}{2} \ln \left\{\operatorname{Det}\left(\beta\left[G_0^{+}\right]_{ub}^{-1}\right\}\left\{\operatorname{Det}\left(\beta\left[G_0^{-}\right]_{ub}^{-1}\right)\right\}\right. 
 \times\left\{\operatorname{Det}\left(\beta\left[G_0^{+}\right]_{db}^{-1}\right) \operatorname{Det}\left(\beta\left[G_0^{-}\right]_{db}^{-1}\right)\right\}\ . \label{blue partition F}
\end{aligned}
\end{equation}
We first work out
\begin{equation}
\begin{aligned}
& \left\{\operatorname{Det}\left(\beta\left[G_0^{+}\right]_{ub}^{-1}\right)\operatorname{Det}\left(\beta\left[G_0^{-}\right]_{ub}^{-1}\right)\right\}
=\beta^4\left[p_0^2-E_{ub}^{+^2}\right]\left[p_0^2-E_{ub}^{-2}\right]\ , \\
& \left\{\operatorname{Det}\left(\beta\left[G_0^{+}\right]_{db}^{-1}\right)\operatorname{Det}\left(\beta\left[G_0^{-}\right]_{db}^{-1}\right)\right\}
=\beta^4\left[p_0^2-E_{db}^{+^2}\right]\left[p_0^2-E_{db}^{-2}\right]\ ,
\end{aligned}
\end{equation}
with $E_{ub}^{\pm}=E_{\boldsymbol{p}} \pm \mu_{ub}$, $E_{db}^{\pm}=E_{\boldsymbol{p}} \pm \mu_{db}$ and $E_{\boldsymbol{p}}=\sqrt{\boldsymbol{p}^2+M^2}$. 

Considering the determinant in the flavor, color, spin spaces, and momentum-frequency space, the expression of Eq.~(\ref{blue partition F}) can be obtained as follows
\begin{equation}
\begin{aligned}
\ln \mathcal{Z}_{b}= & \sum_n \sum_{\vec{p}}\left\{\ln \left(\beta^2\left[p_0^2-E_{ub}^{+^2}\right]\right)+\ln \left(\beta^2\left[p_0^2-E_{ub}^{-2}\right]\right)\right. \\
& \left.+\ln \left(\beta^2\left[p_0^2-E_{db}^{+^2}\right]\right)+\ln \left(\beta^2\left[p_0^2-E_{db}^{-2}\right]\right)\right\} .
\end{aligned}
\end{equation}

\subsection{Calculating the red and green quarks' partition function under charge-neutrality conditions} \label{charge neutral}

Before we evaluate the partition function for the quarks participating in the diquark condensate, we will first present some model-independent results.
The partition function in the momentum space has the following relation 
\begin{equation}
\begin{aligned}
\ln \mathcal{Z}=\frac{1}{2} \ln \operatorname{Det}\left(\beta G^{-1}\right)\ . \label{Appendix eq partition}
\end{aligned}
\end{equation}
For a $2 \times 2$ matrix with elements $A, B, C$, and $D$, we have the identity
\begin{equation}
\begin{aligned}
\operatorname{Det}\left(\begin{array}{ll}
A & B \\
C & D
\end{array}\right) & =\operatorname{Det}\left(-C B+C A C^{-1} D\right) 
 =\operatorname{Det}\left(-B C+D C^{-1} A C\right) \ .
\end{aligned}
\end{equation}
To prove the above equation, we have used
\begin{equation}
\begin{aligned}
\left(\begin{array}{ll}
A & B \\
C & D
\end{array}\right) & \equiv\left(\begin{array}{cc}
0 & B \\
C & 0
\end{array}\right)\left(\begin{array}{cc}
1 & C^{-1} D \\
B^{-1} A & 1
\end{array}\right) 
 \equiv\left(\begin{array}{ll}
B C^{-1} & A B^{-1} \\
D C^{-1} & C B^{-1}
\end{array}\right)\left(\begin{array}{cc}
0 & C \\
B & 0
\end{array}\right) \ .
\end{aligned}
\end{equation}
Replacing $A, B, C$, and $D$ with the corresponding elements of $G^{-1}$, we have
\begin{equation}
\begin{aligned}
\operatorname{Det}\left(\beta \mathrm{G}^{-1}\right)= & \beta^2 \operatorname{Det} D_1=\beta^2 \operatorname{Det}\left(-\Delta^{+} \Delta^{-}\right. 
\left.+\Delta^{+}\left[G_0^{+}\right]^{-1}\left[\Delta^{+}\right]^{-1}\left[G_0^{-}\right]^{-1}\right) \\
= & \beta^2 \operatorname{Det} D_2=\beta^2 \operatorname{Det}\left(-\Delta^{+} \Delta^{-}\right.  \left.+\left[G_0^{-}\right]^{-1}\left[\Delta^{-}\right]^{-1}\left[G_0^{+}\right]^{-1} \Delta^{-}\right) \ .
\end{aligned}
\end{equation}
Using the energy projectors $\tilde{\Lambda}_{\pm}$, we can work out $D_1$ and $D_2$ as
\begin{equation}
\begin{aligned}
D_1= & \Delta^2+\gamma_5\left[\gamma_0\left(p_0-E_{\boldsymbol{p}}^{-}\right) \tilde{\Lambda}_{+}+\gamma_0\left(p_0+E_{\boldsymbol{p}}^{+}\right) \tilde{\Lambda}_{-}\right] 
 \times \gamma_5\left[\gamma_0\left(p_0-E_{\boldsymbol{p}}^{+}\right) \tilde{\Lambda}_{+}+\gamma_0\left(p_0+E_{\boldsymbol{p}}^{-}\right) \tilde{\Lambda}_{-}\right] \\
= & -\left[\left(p_0^2-E_{\boldsymbol{p}}^{-2}-\Delta^2\right) \tilde{\Lambda}_{-}+\left(p_0^2-E_{\boldsymbol{p}}^{+2}-\Delta^2\right) \tilde{\Lambda}_{+}\right] \\
D_2= & -\left[\left(p_0^2-E_{\boldsymbol{p}}^{-2}-\Delta^2\right) \tilde{\Lambda}_{+}+\left(p_0^2-E_{\boldsymbol{p}}^{+2}-\Delta^2\right) \tilde{\Lambda}_{-}\right] \ .\label{D1D2}
\end{aligned}
\end{equation}
Using the properties of the energy projectors, we can get
\begin{equation}
\begin{aligned}
D_1 D_2=&\left[p_0^2-(E_{\boldsymbol{p}}^-)^2-\Delta^{2}\right]\left[p_0^2-(E_{\boldsymbol{p}}^+)^2-\Delta^{2}\right]
=\left[p_0^2-E_{\Delta}^{-2}\right]\left[p_0^2-E_{\Delta}^{+2}\right]\ .
\end{aligned}
\end{equation}
Then Eq.~(\ref{Appendix eq partition}) can be derived as
\begin{equation}
\begin{aligned}
\ln \mathcal{Z} & =\frac{1}{2} \ln \operatorname{Det}\left(\beta G^{-1}\right)=\frac{1}{4} \operatorname{Tr} \ln \left[\beta^2 D_1 \beta^2 D_2\right] \\
& =\frac{1}{4}\left\{\operatorname{Tr} \ln \left[\beta^2\left(p_0^2-E_{\Delta}^{-2}\right)\right]+\operatorname{Tr} \ln \left[\beta^2\left(p_0^2-E_{\Delta}^{+2}\right)\right]\right\}\ .
\end{aligned}
\end{equation}

Under charge-neutrality conditions, the interacting particles have different chemical potentials. From Eqs.~(\ref{mean potential}) and (\ref{difference Potential}), we have
\begin{equation}
\begin{aligned}
\tilde{\mu}_{ur}& =\bar{\mu}- \delta \mu\ , \\
\tilde{\mu}_{dg} & =\bar{\mu}+ \delta \mu\ ,
\end{aligned}
\end{equation}
The propagator of $q_{ur}$ paired with $q_{dg}$ is 
\begin{equation}
\begin{aligned}
G_{ur,dg}^{-1}=\left(\begin{array}{cc}
\gamma^{\mu}p_{\mu}+\gamma^0 \tilde{\mu}_{ur}-M & \Delta^{-} \\
\Delta^{+} & \gamma^{\mu}p_{\mu}- \gamma^0 \tilde{\mu}_{dg}-M
\end{array}\right) \ .
\end{aligned}
\end{equation}
%
Following similar calculations as in Eq.~(\ref{D1D2}), we can find that
\begin{equation}
\begin{aligned}
\operatorname{Det} G_{ur,dg}^{-1}&= \operatorname{Det} D_1
\end{aligned}
\end{equation}
with
\begin{equation}
\begin{aligned}
D_1&=  \Delta^2 +\gamma_5\left[\gamma_0\left(p_0-\left(E_{\boldsymbol{p}}-\tilde{\mu}_{ur}\right)\right) \tilde{\Lambda}_{+}+\gamma_0\left(p_0+\left(E_{\boldsymbol{p}}+\tilde{\mu}_{ur}\right)\right) \tilde{\Lambda}_{-}\right] \\
& \times \gamma_5\left[\gamma_0\left(p_0-\left(E_{\boldsymbol{p}}+\tilde{\mu}_{dg}\right)\right) \tilde{\Lambda}_{+}+\gamma_0\left(p_0+\left(E_{\boldsymbol{p}}-\tilde{\mu}_{dg}\right)\right) \tilde{\Lambda}_{-}\right] \\
&=  \Delta^2 +\gamma_5\left[\gamma_0\left(\left(p_0-\delta \mu\right)-\left(E_{\boldsymbol{p}}-\bar{\mu}\right)\right)\tilde{\Lambda}_{+} \right.
\left.+\gamma_0\left(\left(p_0-\delta \mu\right)+\left(E_{\boldsymbol{p}}+\bar{\mu}\right)\right) \tilde{\Lambda}_{-}\right] \\
& \times \gamma_5\left[\gamma_0\left(\left(p_0-\delta \mu\right)-\left(E_{\boldsymbol{p}}+\bar{\mu}\right)\right) \tilde{\Lambda}_{+}\right.
\left.+\gamma_0\left(\left(p_0-\delta \mu\right)+\left(E_{\boldsymbol{p}}-\bar{\mu}\right)\right) \tilde{\Lambda}_{-}\right] \\
&= -\left[\left(\left(p_0 -\delta \mu\right)^2-E_{\boldsymbol{p}}^{-2}-\Delta^2\right) \tilde{\Lambda}_{-}\right.
\left. +\left(\left(p_0-\delta \mu\right)^2-E_{\boldsymbol{p}}^{+2}-\Delta^2\right) \tilde{\Lambda}_{+}\right]\ ,
\end{aligned}
\end{equation}
and 
\begin{equation}
\begin{aligned}
D_2 &= 
-\left[\left(\left(p_0 +\delta \mu\right)^2-E_{\boldsymbol{p}}^{-2}-\Delta^2\right) \tilde{\Lambda}_{+}\right.
\left.+\left(\left(p_0+\delta \mu\right)^2-E_{\boldsymbol{p}}^{+2}-\Delta^2\right) \tilde{\Lambda}_{-}\right] \ .
\end{aligned}
\end{equation}
Following a similar way, we can get the propagator of $q_{dg}$ paired with $q_{ur}$ as
\begin{equation}
\begin{aligned}
G_{dg,ur}^{-1}=\left(\begin{array}{cc}
\gamma^{\mu}p_{\mu}+\gamma^0 \tilde{\mu}_{dg}-M & \Delta^{-} \\
\Delta^{+} & \gamma^{\mu}p_{\mu}- \gamma^0 \tilde{\mu}_{ur}-M
\end{array}\right)\ ,
\end{aligned}
\end{equation}
and we further obtain
\begin{equation}
\begin{aligned}
\operatorname{Det} G_{dg,ur}^{-1}&= \operatorname{Det} D_3\ ,
\end{aligned}
\end{equation}
in which 
\begin{equation}
\begin{aligned}
D_3&= -\left[\left(\left(p_0+\delta \mu\right)^2-E_{\boldsymbol{p}}^{-2}-\Delta^2\right) \tilde{\Lambda}_{+}\right.
\left. +\left(\left(p_0+\delta \mu\right)^2-E_{\boldsymbol{p}}^{+{ }^2}-\Delta^2\right) \tilde{\Lambda}_{-}\right]\ ,\\
D_4& =-\left[\left(\left(p_0-\delta \mu\right)^2-E_{\boldsymbol{p}}^{-2}-\Delta^2\right) \tilde{\Lambda}_{-}\right.
\left.+\left(\left(p_0-\delta \mu\right)^2-E_{\boldsymbol{p}}^{+2}-\Delta^2\right) \tilde{\Lambda}_{+}\right]  \ .
\end{aligned}
\end{equation}
Then the partition function of red and green quarks that participate in pairing can be obtained as follows:
\begin{equation}
\begin{aligned}
\ln \mathcal{Z}_{r,g}= & \frac{1}{2} \ln \operatorname{Det}\left(\beta G^{-1}\right) \\
= &\frac{1}{4}\left\{\operatorname{Tr} \ln \left[\beta^2 D_1 \beta^2 D_2\right]+\operatorname{Tr} \ln \left[\beta^2 D_3 \beta^2 D_4\right]\right\} \\
= & \frac{1}{4}\left\{\operatorname{Tr} \ln \left[\beta^2\left(p_0^2-\left(E_{\Delta}^{-} +\delta \mu\right)^2\right)\right] \right.
\left.+\operatorname{Tr} \ln \left[\beta^2\left(p_0^2-\left(E_{\Delta}^{+}+\delta \mu\right)^2\right)\right]\right.\\ 
& \left.+\operatorname{Tr} \ln \left[\beta^2\left(p_0^2-\left(E_{\Delta}^{-}-\delta \mu\right)^2\right)\right]\right.
\left.+\operatorname{Tr} \ln \left[\beta^2\left(p_0^2-\left(E_{\Delta}^{+}-\delta \mu\right)^2\right)\right]\right\}\ .
\end{aligned}
\end{equation}

\section{Evaluating the thermodynamic potential}
After deriving the $\mathcal{Z}_{b}$ unpairing part and $\mathcal{Z}_{r,g}$ pairing part in Eq.~(\ref{partition function}), we can finally evaluate the thermodynamic potential~\citep{2011ftft.book.....K}. The frequency summation of the free energy
\begin{equation}
\begin{aligned}
\ln \mathcal{Z}_f=\sum_n \ln \left[\beta^2\left(p_0^2-E_{\boldsymbol{p}}^2\right)\right] \label{free energy}
\end{aligned}
\end{equation}
can always be obtained by performing the frequency summation of the propagator $1 /\left(p_0^2-E_{\boldsymbol{p}}^2\right)$. 
By differentiating Eq.~(\ref{free energy}) with respect to $E_{\boldsymbol{p}}$ :
\begin{equation}
\begin{aligned}
\frac{\partial \ln \mathcal{Z}_f}{\partial E_p}=-2 E_{\boldsymbol{p}} \sum_n \frac{1}{p_0^2-E_p^2}=\beta\left[1-2 \tilde{f}\left(E_{\boldsymbol{p}}\right)\right]\ ,
\end{aligned}
\end{equation}
where $\tilde{f}(x)=1 /\left(e^{\beta x}+1\right)$ is the usual Fermi-Dirac distribution function,
then integrating with respect to $E_{\boldsymbol{p}}$, one can get the free energy
\begin{equation}
\begin{aligned}
\ln \mathcal{Z}_f=\beta\left[E_{\boldsymbol{p}}+2 T \ln \left(1+e^{-\beta E_{\boldsymbol{p}}}\right)\right]\ .
\end{aligned}
\end{equation}
With the help of the above expression, and by replacing
\begin{equation}
\begin{aligned}
\sum_p \rightarrow V \int \frac{d^3 \boldsymbol{p}}{(2 \pi)^3}\ ,
\end{aligned}
\end{equation}
we can obtain the thermodynamic potential of the quark-matter system as written in Eq.~(\ref{therm.}):
\begin{equation}
\begin{aligned}
\Omega_q &=  \frac{\sigma^{2}}{4 G_S}-\frac{\left(\mu-\tilde{\mu}\right)^2}{4 G_V}+\frac{\Delta^2}{4 G_D}
-2 \int \frac{d^3 \boldsymbol{p}}{(2 \pi)^3}\left\{2 E_p+2 E_{\Delta}^{+}+2 E_{\Delta}^{-}\right. \\
& +T \ln \left[1+\exp \left(-\beta E_{ub}^{+}\right)\right]+T \ln \left[1+\exp \left(-\beta E_{ub}^{-}\right)\right] \\
& +T \ln \left[1+\exp \left(-\beta E_{db}^{+}\right)\right]+T \ln \left[1+\exp \left(-\beta E_{db}^{-}\right)\right] \\
& +2 T \ln \left[1+\exp \left(-\beta E_{\Delta^{+}}^{+}\right)\right]+2 T \ln \left[1+\exp \left(-\beta E_{\Delta^{-}}^{+}\right)\right] \\
& \left.+2 T \ln \left[1+\exp \left(-\beta E_{\Delta^{+}}^{-}\right)\right]+2 T \ln \left[1+\exp \left(-\beta E_{\Delta^{-}}^{-}\right)\right]\right\}\ .
\end{aligned}
\end{equation}

\section*{Acknowledgements}
We appreciate the valuable discussions with Wen-Hao Zhou and Li-Qun Su in the numerical calculations. We also thank Bo-Lin Li, Chen Zhang, Mei Huang, Yan Yan, Yi-Lun Du, and the XMU neutron star group for helpful discussions.
The work is supported by the National SKA Program of China (No.~2020SKA0120300), the National Natural Science Foundation of China (Grant No. 12273028), and Jiangsu Province Postgraduate Research and Practice Innovation Programme (Grant no. KYCX23$\_$0100).


\end{document}